\documentstyle[12pt,epsf]{article}
\renewcommand{\baselinestretch}{1.30}
\begin{document}
\begin{flushright}
February, 1999
\end{flushright}
\vspace{0mm}
\begin{center}
\large{On the Chiral Quark Soliton Model with Pauli-Villars Regularization}
\end{center}
\vspace{0mm}
\begin{center}
T.~Kubota\footnote{Email \ : \ kubota@kern.phys.sci.osaka-u.ac.jp}
and  M.~Wakamatsu\footnote{Email \ : \ wakamatu@miho.rcnp.osaka-u.ac.jp}
\end{center}
\vspace{-4mm}
\begin{center}
Department of Physics, Faculty of Science, \\
Osaka University, \\
Toyonaka, Osaka 560, JAPAN
\end{center}
\vspace{0mm}
\begin{center}
T.~Watabe\footnote{Email \ : \ watabe@rcnp.osaka-u.ac.jp}
\end{center}
\vspace{-4mm}
\begin{center}
Research Center for Nuclear Physics (RCNP), \\
Osaka University, \\
Ibaraki, Osaka 567, JAPAN
\end{center}

\vspace{8mm}

\vspace{10mm}
\begin{center}
\small{{\bf Abstract}}
\end{center}
\vspace{-1mm}
\begin{center}
\begin{minipage}{15.5cm}
\renewcommand{\baselinestretch}{1.0}
\small
The Pauli-Villars regularization scheme is often used
for evaluating parton distributions within the framework of the
chiral quark soliton model with inclusion of the vacuum polarization
effects. Its simplest version with a single subtraction term should
however be taken with some caution, since it does not fully get rid
of divergences contained in scalar and psuedoscalar quark densities
appearing in the soliton equation of motion. To remedy this shortcoming,
we propose here its natural extention, i.e. the Pauli-Villars
regularization scheme with multi-subtraction terms. We also carry out
a comparative analysis of the Pauli-Villars regularization scheme
and more popular proper-time one. It turns out that some isovector
observables like the isovector magnetic moment of the nucleon is rather
sensitive to the choice of the regularization scheme.
In the process of tracing the origin of this sensitivity, a noticeable
difference of the two regularization scheme is revealed.

\normalsize
\end{minipage}
\end{center}
\renewcommand{\baselinestretch}{2.0}

\newpage
\section{Introduction}

The recent calculations of nucleon parton distributions within
the chiral quark soliton model (CQSM) exclusively utilizes the so-called
Pauli-Villars regularization scheme [1-6]. This is to be contrasted with the
fact that most of the past calculations of the nucleon static observables
were carried out by using the proper-time regularization scheme [7-8].
There are some reasons for it. The first reason is mainly technical.
For obtaining parton distributions, one need to evaluate nucleon matrix
elements of quark bilinear operators which is nonlocal in two
space-time coordinates. The problem is that we have no unanimous idea
about how to generalize the proper-time scheme for the reguralization of
such unusual quantities. The second but more positive reason for using the
Pauli-Villars regularization scheme has been advocated by Diakonov
et al. [1,2].
They emphasize that this regularization scheme preserves certain general
properties of parton distributions such as positivity, factorization
properties, sum rules etc., which are easily violated by other regularization
schemes like the proper-time one.

Recently, there was a controversial debate on the stability of
soliton solutions in the CQSM regularized with the Pauli-Villars
subtraction scheme [10,11].
It seems that the problem has been settled by now,
since stable soliton solutions seem to exist at any rate if the
Pauli-Villars regularization is applied to the quark seas only, not to
the discrete bound state sometimes called the valence
quark orbital. Unfortunately, this is not the end of the story.
In fact, soliton solutions of the CQSM with use of the Pauli-Villars
regularization scheme were obtained many years ago by D\"{o}ring
et al. [12]. (To be more precise, the model used by them is not the
CQSM but the Nambu-Jona-Lasinio model. In fact, they were forced to
impose an {\it ad hoc} nonlinear constraint for the scalar and
pseudoscalar meson fields at the later stage of manipulation.
Otherwise, they would not have obtained any convergent solutions [13].)
The fact that the single-subtraction Pauli-Villars scheme cannot
regularize the vacuum quark condensate was already noticed in
an earlier paper [14] as well as in this paper [12].
To remove this divergence, which is necessary for obtaining a finite
gap equation, D\"{o}ring et al. propose to add some counter terms,
which depend on the meson fields, to the original effective action.
It is very important to
recognize that this procedure is not workable within the CQSM, since
their counter terms reduce to mere constants under the chiral circle
condition  which we impose from the very beginning. Thus, one must
conclude that the simplest Pauli-Villars scheme with the
single-subtraction term is unable to fully get rid of the divergence of
the vacuum quark condensate at least in the nonlinear model.
One should take this fact seriously, because
it brings about a trouble also in the physics of soliton sector.
To understand it, one has only to remember the
fact that the scalar quark density appearing in the soliton equation
of motion is expected to approach a finite and nonzero value
characterizing the vacuum quark condensate as the distance from
the soliton center becomes large [15].
This necessarily means that the
scalar quark density appearing in the soliton equation of motion
cannot also be free from divergences.

The purpose of the present study is then twofold. On the one hand,
we want to show that the single-subtraction Pauli-Villars scheme is
not a fully satisfactory regularization scheme, and that at least one more
subtraction term is necessary for a consistent regularization
of the effective theory. This will be made convinced through the
formal discussion given in II and also the explicit numerical
results shown in III.A. On the other hand, we also want to know
the regularization-scheme dependence of the CQSM through the comparative
analysis of typical static observables of the nucleon predicted by the two
regularization schemes, i.e. the Pauli-Villars one and the
proper-time one. The discussion on this second issue will
be given in III.B. We then summarize our conclusion in IV.

\vspace{4mm}
\section{Pauli-Villars regularization scheme}

We begin with the effective lagrangian of the chiral quark model
with an explicit chiral symmetry breaking term as
\begin{equation}
   {\cal L}_{CQM} = {\cal L}_0 + {\cal L}^\prime ,
\end{equation}
where ${\cal L}_0$ denotes the chiral symmetric part [16] given by
\begin{eqnarray}
   {\cal L}_0 = \bar{\psi} \,(\,i \not\!\partial - 
   M U^{\gamma_5} (x) \,) \psi ,
\end{eqnarray}
with
\begin{equation}
  U^{\gamma_5} (x) = e^{\,i \gamma_5 \mbox{\boldmath $\tau$}
  \cdot \mbox{\boldmath $\pi$} (x) / f_\pi \,} = 
  \frac{1 + \gamma_5}{2} \,U(x) + \frac{1 - \gamma_5}{2} \,U^\dagger (x),
\end{equation}
while

\begin{equation}
   {\cal L}^\prime = \frac{1}{4} f_\pi^2 m_\pi^2 \,\mbox{tr}
   ( U(x) + U^\dagger (x) - 2),
\end{equation}
is thought to simulate a small deviation from the chiral symmetric limit.
Here the trace in (4) is to be taken with respect to flavor indices.
(One could have taken an alternative choice that introduces explicit
chiral-symmetry-breaking effects in the form of quark mass term.
We did not do so, because it turns out that this form of action
cannot be regularized consistently with the Pauli-Villars
subtraction method.)

The idea of the Pauli-Villars regularization can most easily be
understood by examining the form of the effective meson action
derived from (1) with the help of the standard derivative expansion :
\begin{equation}
   S_{eff} [U] = S_f [U] + S_m [U] ,
\end{equation}
where
\begin{eqnarray}
   S_f [U] &=& - \,i \,N_c \,\mbox{Sp} \log ( i \! \not\!\partial - 
   M U^{\gamma_5} ) \nonumber \\
   &=& \int d^4 x \,\{ 4 N_c M^2 I_2 (M) 
   \,\mbox{tr} ( \partial_\mu U \partial^\mu U^\dagger ) + 
   \mbox{higher derivative terms} \} ,\\
   S_m [U] &=& \int d^4 x \,\frac{1}{4} f_\pi^2 m_\pi^2
   \,\mbox{tr} ( U(x) + U^\dagger (x) - 2) .
\end{eqnarray}
In eq.(6), the coefficient
\begin{equation}
   I_2 (M) \equiv - i \int \frac{d^4 k}{{(2 \pi)}^4} 
   \frac{1}{{(k^2 - M^2)}^2} ,
\end{equation}
of the pion kinetic term diverges logarithmically. In fact, by
introducing a ultraviolet cutoff momentum $\alpha$ that should eventually
be made infinity, one finds that
\begin{equation}
   I_2 (M) \sim \frac{1}{16 \pi^2} \{ \ln \alpha^2 - \ln M^2 - 1 \} .
\end{equation}
This logarithmic divergence can be removed if one introduces a
regularized action as follows :
\begin{equation}
   S_{eff}^{reg} [U] = S_f^{reg} [U] + S_m [U] ,
\end{equation}
where
\begin{equation}
   S_f^{reg} [U] \equiv S_f [U] - {\left( \frac{M}{M_{PV}} \right)}^2
   S_f^{M_{PV}} [U] .
\end{equation}
Here $S_f^{M_{PV}}$ is obtained from $S_f [U]$ with $M$ replaced by
the Pauli-Villars regulator mass $M_{PV}$. Further requiring that the
above regularized action reproduces correct normalization for the pion
kinetic term, one obtains the condition :
\begin{equation}
   \frac{N_c M^2}{4 \pi^2} \ln {\left( \frac{M_{PV}}{M} \right)}^2 = 
   f_\pi^2 ,
\end{equation}
which can be used to fix the regulator mass $M_{PV}$. Once the
effective action is regularized, the static soliton energy should be a
finite functional of the soliton profile $F(r)$ under the standard
hedgehog ansatz $U (\mbox{\boldmath $x$}) = 
\exp [ i \mbox{\boldmath $\tau$}
\cdot \hat{\mbox{\boldmath $r$}} F(r) ]$. Since the soliton equation
of motion is obtained from the stationary condition of the static
energy against the variation of $F(r)$, everything seems to be going
well with the above single-subtraction Pauli-Villars regularization
procedure. Unfortunately, this is not the case. To understand what the
problem is, we first recall the fact that the scalar quark density
appearing in the soliton equation of motion is expected to approach a
finite and nonzero constant characterizing the vacuum quark condensate
as the distance from the soliton center becomes large [15].
(This is a natural consequence of our demand that both of the soliton
($B=1$) and vacuum($B=0$) sectors must be described by the same (or
single) equation of motion.) On the other hand,
it has been known that the vacuum quark condensate contains quadratic
divergences that cannot be removed by the single-subtraction
Pauli-Villars scheme [12,14]. This then indicates that the scalar quark
density appearing in the soliton equation of motion cannot also be free
from divergences.

To get rid of all the troublesome divergences, we propose here to
increase the number of subtraction terms, thereby starting with the
following action :
\begin{equation}
   S_{eff}^{reg} [U] = S_f^{reg} [U] + S_m [U] ,
\end{equation}
where
\begin{equation}
   S_f^{reg} [U] \equiv S_f [U] - \sum_{i = 1}^{N}
   c_i S_f^{\Lambda_i} [U] ,
\end{equation}
with $N$ being the number of subtraction terms. The logarithmic divergence
of the original action is removed if the condition
\begin{equation}
   1 - \sum_{i = 1}^{N} c_i {\left( \frac{\Lambda_i}{M} \right)}^2
   = 0
\end{equation}
is fulfilled. Similarly, the normalization condition (12) is
replaced by
\begin{equation}
   \frac{N_c M^2}{4 \pi^2} \sum_{i = 1}^N c_i
   {\left( \frac{\Lambda_i}{M} \right)}^2 \ln 
   {\left( \frac{\Lambda_i}{M} \right)}^2 = f_\pi^2 .
\end{equation}
The single-subtraction Pauli-Villars scheme corresponds to taking
$N = 1, \Lambda_1 = M_{PV}$, and $c_1 = {( M / M_{PV} )}^2$.
This is naturally the simplest case that satisfies both conditions
(15) and (16).

To derive soliton equation of motion, we must first write down a
regularized expression for the static soliton energy.
Under the hedgehog ansatz $\mbox{\boldmath $\pi$} 
(\mbox{\boldmath $x$}) = f_\pi \hat{\mbox{\boldmath $r$}} F(r)$ for
the background pion fields, it is obtained in the form :
\begin{equation}
   E_{static}^{reg} [F(r)] = E_f^{reg} [F(r)] + E_m [F(r)] ,
\end{equation}
where the meson part is given by
\begin{equation}
   E_m [F(r)] = - f_\pi^2 m_\pi^2 \int d^3 x \left(
   \cos F(r) - 1 \right) ,
\end{equation}
while the fermion (quark) part is given as
\begin{equation}
   E_f^{reg} [F(r)] = E_{val} + E_{vp}^{reg} ,
\end{equation}
with
\begin{eqnarray}
   E_{val} &=& N_c E_0 \\
   E_{vp}^{reg} &=& N_c \sum_{n < 0} \left( E_n - E_n^{(0)} \right) - 
   \sum_{i = 1}^N c_i \,N_c \sum_{n < 0} \left( E_n^{\Lambda_i} - 
   E_n^{(0) \Lambda_i} \right) .
\end{eqnarray}
Here $E_n$ are the quark single-particle energies, given as the
eigenvalues of the static Dirac hamiltonian in the background pion
fields :
\begin{equation}
   H \,| \,n > = E_n \,| \,n > ,
\end{equation}
with
\begin{equation}
   H = \frac{\mbox{\boldmath $\alpha$} \cdot \nabla}{i} 
   + \beta M
   \left( \cos F(r) + i 
   \gamma_5 \mbox{\boldmath $\tau$} \cdot 
   \hat{\mbox{\boldmath $r$}} \sin F(r) \right) ,
\end{equation}
while the energy $E_n^{(0)}$ denote the energy eigenvalues of the
vacuum hamiltonian given by eq.(23) with $F(r) = 0$ or $U = 1$.
Eq.(19) means that the quark part of the static energy is given as
a sum of the contribution of the discrete bound-state level and
that of the negative energy Dirac continuum. The latter part
is regularized by subtracting from the Dirac sea contribution a
linear combination of the corresponding sum evaluated with the
regulator mass $\Lambda_i$ instead of the dynamical quark mass.
($E_n^{\Lambda_i}$ in these subtraction terms are the eigenenergies
of the Dirac hamiltonian (23) with $M$ replaced by $\Lambda_i$
and with the same background pion field.)

Now the soliton equation of motion is obtained from the stationary
condition of $E_{static}^{reg} [F(r)]$ with respect to the variation
of the profile function $F(r)$ :
\begin{eqnarray}
   0 &=& \frac{\delta E_{static} [F(r)]}{\delta F(r)} \nonumber \\
   &=& 4 \pi r^2 \left\{ - M \left[ 
   S(r) \sin F(r) - P(r) \cos F(r) \right] + 
   f_\pi^2 m_\pi^2 \sin F(r) \right\} ,
\end{eqnarray}
which gives
\begin{equation}
   F(r) = \arctan \left( \frac{P(r)}{S(r) - 
   \frac{f_\pi^2 m_\pi^2}{M}} \right) .
\end{equation}
Here $S(r)$ and $P(r)$ are regularized scalar and pseudoscalar
densities given as
\begin{eqnarray}
   S(r) &=& S_{val} (r) + \sum_{n < 0} S_n (r) - 
   \sum_{i = 1}^N c_i \frac{\Lambda_i}{M} \sum_{n < 0} 
   S_n^{\Lambda_i} (r) , \\
   P(r) &=& P_{val} (r) + \sum_{n < 0} P_n (r) - 
   \sum_{i = 1}^N c_i \frac{\Lambda_i}{M} \sum_{n < 0} 
   P_n^{\Lambda_i} (r) ,   
\end{eqnarray}
with
\begin{eqnarray}
   S_n (r) &=& \frac{N_c}{4 \pi} \int d^3 x 
   < n | \mbox{\boldmath $x$} > \,\gamma^0 \,
   \frac{\delta ( | \mbox{\boldmath $x$} | - r )}{r^2}
   < \mbox{\boldmath $x$} | n > , \\
   P_n (r) &=& \frac{N_c}{4 \pi} \int d^3 x 
   < n | \mbox{\boldmath $x$} > \,i \gamma^0 \gamma_5 \,
   \mbox{\boldmath $\tau$} \cdot \hat{\mbox{\boldmath $r$}} \,
   \frac{\delta ( | \mbox{\boldmath $x$} | - r )}{r^2}
   < \mbox{\boldmath $x$} | n > ,
\end{eqnarray}
and $S_{val} (r) = S_{n = 0} (r)$ and $P_{val} (r) = P_{n = 0} (r)$,
while $S_n^{\Lambda_i} (r)$ and $P_n^{\Lambda_i} (r)$ are the
corresponding densities evaluated with the regulator mass $\Lambda_i$
instead of the dynamical quark mass $M$.
As usual, a self-consistent soliton solution is obtained in an
iterative way. First by assuming an appropriate (though arbitrary)
soliton profile $F(r)$, the eigenvalue problem of the Dirac hamiltonian
is solved. Using the resultant eigenfunctions and their associated
eigenenergies, one can calculate the regularized scalar and
pseudoscalar quark densities $S(r)$ and $P(r)$. Eq.(25) can then be
used to obtain a new soliton profile $F(r)$. The whole procedure above
is repeated with this new profile $F(r)$ until the self-consistency
is fulfilled.

Now we recall an important observation made before. The scalar quark density
$S (r)$ at the spatial infinity $r = \infty$ with respect to the
soliton center should coincide with the scalar quark density in the
vacuum ($B = 0$) sector, which is nothing but the familiar vacuum
quark condensate (per unit volume) ${\langle \bar{\psi} 
\psi \rangle}_{vac}$. That is, the following simple relation must hold :
\begin{equation}
   {\langle \bar{\psi} \psi \rangle}_{vac} \ = \ \frac{1}{V} 
   \int S (r = \infty) \,d^3 r \ = \ S(r = \infty) .
\end{equation}
(Later, this relation will be checked numerically.)
What we must do now is to find necessary conditions for the subtraction
constants $c_i$ and $\Lambda_i$ in the multi-subtraction Pauli-Villars
scheme to make the vacuum quark condensate finite.
This can be achieved by examining the expression of the vacuum quark
condensate obtained consistently with the soliton
equation of motion :
\begin{equation}
   M {\langle \bar{\psi} \psi \rangle}_{vac}^{reg} = 
   M {\langle \bar{\psi} \psi \rangle}_{vac} - \sum_{i = 1}^N c_i
   \left( \frac{\Lambda_i}{M} \right) \Lambda_i 
   {\langle \bar{\psi} \psi \rangle}_{vac}^{\Lambda_i} ,
\end{equation}
or equivalently
\begin{equation}
   {\langle \bar{\psi} \psi \rangle}_{vac}^{reg} = 
   {\langle \bar{\psi} \psi \rangle}_{vac} - \sum_{i = 1}^N c_i
   {\left( \frac{\Lambda_i}{M} \right)}^2
   {\langle \bar{\psi} \psi \rangle}_{vac}^{\Lambda_i} ,
\end{equation}
where
\begin{equation}
   {\langle \bar{\psi} \psi \rangle}_{vac} = - 4 N_c M 
   \int \frac{d^3 k}{{(2 \pi)}^3} \frac{1}{E_k^{(0)}} ,
\end{equation}
with $E_k^{(0)} = {( k^2 + M^2)}^{1 / 2}$, while
${\langle \bar{\psi} \psi \rangle}_{vac}^{\Lambda_i}$ are obtained
from ${\langle \bar{\psi} \psi \rangle}_{vac}$ with the replacement
of $M$ by $\Lambda_i$. Using the integration formula
\begin{equation}
   \int^\alpha \frac{d^3 k}{{(2 \pi)}^3} \frac{1}{\sqrt{k^2 + M^2}} = 
   \frac{1}{8 \pi^2} \left\{ 2 \alpha^2 - M^2 \ln \alpha^2 +
   (1 - 2 \ln 2) M^2 + M^2 \ln M^2 \right\} ,
\end{equation}
with $\alpha$ being a ultraviolet cutoff momentum, we obtain
\begin{eqnarray}
   {\langle \bar{\psi} \psi \rangle }_{vac}^{reg} &=&
   - \frac{N_c M}{2 \pi^2} 
   \Biggl\{ \left[ 1 - \sum_{i = 1}^N c_i 
   {\left( \frac{\Lambda_i}{M} \right)}^2 \right] \cdot 2 \alpha^2 
   \ - \ \left[ M^2 - \sum_{i = 1}^N c_i
   {\left( \frac{\Lambda_i}{M} \right)}^2 \Lambda_i^2 \right]
   \cdot \ln \alpha^2 \nonumber \\
   &+& \left[ M^2 - \sum_{i = 1}^N c_i
   {\left( \frac{\Lambda_i}{M} \right)}^2 \Lambda_i^2 \right]
   \cdot ( 1 - 2 \ln 2 ) \ + \ M^2 \ln M^2 - \sum_{i = 1}^N c_i
   {\left( \frac{\Lambda_i}{M} \right)}^2 \Lambda_i^2 \ln \Lambda_i^2
   \Biggr\} , \ \ \ \ \ 
\end{eqnarray}
which clearly shows that ${\langle \bar{\psi} \psi \rangle}_{vac}$
contains quadratic and logarithmic divergences as $\alpha$ going to
infinity. These divergences can respectively be removed if the
subtraction constants are chosen to satisfy the following conditions :
\begin{eqnarray}
   M^2 - \sum_{i = 1}^N c_i \Lambda_i^2 &=& 0 , \\
   M^4 - \sum_{i = 1}^N c_i \Lambda_i^4 &=& 0 .
\end{eqnarray}
Using the first of these conditions, the finite part of ${\langle
\bar{\psi} \psi \rangle}_{vac}$ can also be expressed as 
\begin{eqnarray}
   {\langle \bar{\psi} \psi \rangle}_{vac} 
   = \frac{N_c M^3}{2 \pi^2} \sum_{i = 1}^N \,c_i \,
   {\left( \frac{\Lambda_i}{M} \right)}^4 
   \log {\left( \frac{\Lambda_i}{M} \right)}^2 
\end{eqnarray}
It is now obvious that the single-subtraction Pauli-Villars 
scheme cannot satisfy both conditions (36) and (37) simultaneously.
Although the quadratic divergence may be removed, the logarithmic
divergence remains in ${\langle \bar{\psi} \psi \rangle}_{vac}$
and consequently also in $S(r = \infty)$ in view of the relation (30).
 To get rid of both these divergences, we need at least two subtraction
terms, which contains four parameters $c_1,c_2$ and $\Lambda_1, \Lambda_2$.
The strategy for fixing these parameters is as follows. First by solving 
the two equations (36) and (37) with $N = 2$ for $c_1$ and $c_2$, we obtain
\begin{eqnarray}
   c_1 &=& \ \,\,\,{\left(\frac{M}{\Lambda_1}\right)}^2 
   \frac{\Lambda_2^2 - M^2}{\Lambda_2^2 - \Lambda_1^2} , \\
   c_2 &=& - \,{\left(\frac{M}{\Lambda_2}\right)}^2 
   \frac{\Lambda_1^2 - M^2}{\Lambda_2^2 - \Lambda_1^2} ,
\end{eqnarray}
which constrains the values of $c_1$ and $c_2$, once $\Lambda_1$ 
and $\Lambda_2$ are given. For determining $\Lambda_1$ and $\Lambda_2$,
we can then use two conditions (16) and (38), which amounts to adjusting the
normalization of the pion kinetic term and the value of vacuum quark
condensate.

\section{Numerical Results and Discussion}

\subsection{Single- versus double-subtraction Pauli-Villars regularization}

The most important parameter of the CQSM is the dynamical quark
mass $M$, which plays the role of the quark-pion coupling constant thereby
controlling basic soliton properties. Throughout the present investigation,
we use the value $M = 400 \,\mbox{MeV}$ favored from the previous
analyses of static baryon observables.
In the case of single-subtraction Pauli-Villars scheme, the regulator
mass $M_{PV}$ is uniquely fixed to be $M_{PV} = 570.86 \,\mbox{MeV}$ by
using the normalization condition (12) for the pion kinetic term,
and there is no other adjustable parameter in the model. In the case
of double-subtraction Pauli-Villars scheme, we have four regularization
parameters $c_1, c_2, \Lambda_1$, and $\Lambda_2$. From the divergence
free conditions (36) and (37), $c_1$ and $c_2$ are constrained as
(39) and (40), while $\Lambda_1$ and $\Lambda_2$ are determined from
(16) and (38) with $f_\pi = 93 \,\mbox{MeV}$ and
${< \bar{\psi} \psi >}_{vac} = - \,{(286.6 \,\mbox{MeV})}^3$.
In spite of their nonlinearity, the two conditions
(16) and (38) are found to uniquely fix the two parameters $\Lambda_1$ and
$\Lambda_2$ within the physically acceptable range of parameters.
The solution that we found is
\begin{equation}
   c_1 = 0.445, \hspace{6mm} c_2 = -0.00612, \hspace{6mm} 
   \Lambda_1 = 630.01 \,\mbox{MeV}, \hspace{6mm} 
   \Lambda_2 = 1642.13 \, \mbox{MeV}.
\end{equation}
As usual, all the numerical calculations are carried out by using the
so-called Kahana and Ripka basis [17].
Following them, the plane-wave basis, introduced as a set of
eigenstates of the free hamiltonian 
$H_0 = \mbox{\boldmath $\alpha$} \cdot \nabla / i + \beta M$,
is discretized by imposing an appropriate boundary condition
for the radial wave functions at the radius $D$ chosen to be 
sufficiently larger than the soliton size.
The basis is made finite by including only those states with
the momentum $k$ as $k < k_{mox}$. The eigenvalue problem (22)
is then solved by diagonalizing the Dirac hamiltonian $H$ in 
the above basis. We are thus able to solve the self-consistent
Hartree problem and also to calculate any nucleon observables
with full inclusion of the sea-quark degrees of freedom.
If the theory is consistently regularized, final answers
must be stable against increase of $k_{max}$ and $D$
(especially against the increase of $k_{max}$).

Now we show in Fig.1 the $k_{max}$ dependence of the theoretical
pseudoscalar and scalar quark densities in the single-subtraction
Pauli-Villars scheme. These curves are obtained for a fixed value
of $D$ as $MD = 12$. The corresponding $k_{max}$ dependence of
the quark densities in the double-subtraction 
Pauli-Villars scheme are shown in Fig.2.
Comparing the two figures, one immediately notices
that the quark densities obtained in the single-subtraction Pauli-Villars 
scheme do not cease to increase in magnitudes as $k_{max}$ increases.
Undoubtedly, this must be a signal of logarithmic divergences
contained in $S(r = \infty)$ (and generally also in $P(r)$ and $S(r)$).
On the other hand, in the case of double-subtraction Pauli-Villars
scheme, the magnitudes of $P(r)$ and $S(r)$ are seen to grow much more
slowly. To convince more clearly the above qualitative difference of
the two regularization schemes, we plot in Fig.3 the value
of $S(r = \infty)$, i.e. the scalar quark density at the spatial
infinity, as functions of $k_{max}$, and also
as functions of $\log ( k_{max} / M )$.
Contrary to the case of single-subtraction scheme 
in which a clear signal of logarithmic divergence is observed, the
value of $S(r = \infty)$ obtained in the double-subtraction scheme is
seen to converge to some limiting value. Although the rate of this 
convergence is rather slow, it appears that this limiting value
certainly coincides with the prescribed value of vacuum quark
condensate ${\langle \bar{\psi} \psi \rangle}_{vac} = - \,
{(286.6 \,\mbox{MeV})}^3 = - \,3.062 \,\mbox{fm}^{-3}$.

Now that one has convinced the fact that the naive Pauli-Villars
scheme with the single-subtraction term contains logarithmic
divergence in the quark densities appearing in the soliton equation
of motion, one may come to the following question.
Why could the authors of ref.[12] obtain self-consistent soliton solutions
despite the presence of the above-mentioned divergences?
The answer lies in the way of obtaining a self-consistent soliton
profile in the nonlinear model (not in the original 
Nambu-Jona-Lasinio model). After evaluating the pseudoscalar and scalar
quark densities with some (large but) finite model space (especially with
finite $k_{max}$), a new profile function $F(r)$ to be used in the next
iterative step is obtained from (25). Since $P(r)$ and $S(r)$ appears
respectively in the numerator and denominator of the argument of arctangent, 
it can happen that the logarithmic divergence contained 
in both of $P(r)$ and $S(r)$ are offset each other.
(We point out that the effect of the term 
$f_{\pi}^2 m_{\pi}^2 / M$ accompanying the scalar quark density is
rather small, anyway.) In fact, Fig.4 shows the $k_{max}$ dependence
of the self-consistent profile function $F(r)$ in both of the 
single-subtraction scheme and the double-subtraction scheme.
One sees that the resultant $F(r)$ is quite stable against the increase 
of $k_{max}$ even in the single-subtraction scheme, in spite of the fact
that it shows logarithmically divergent behavior for both of $P(r)$ and
$S(r)$. Undoubtedly, this is the reason why the authors of [12] succeeded 
in obtaining self-consistent soliton profile $F(r)$ despite the 
divergences remaining in each of $P(r)$ and $S(r)$.
Because of this fortunate accident, self-consistent 
soliton profiles $F(r)$ in the nonlinear model can be obtained with 
a good accuracy by using a modest value of $k_{max}$ not only for
the double-subtraction scheme but also for the single-subtraction one,
and besides the resultant 
$F(r)$ and not much different in these two schemes. This also 
applies to most nucleon observables which depend only on $F(r)$
and have no direct dependence on $S(r)$ and/or $P(r)$.
The previous calculation of parton distributions with use of 
the single-subtraction Pauli-Villars scheme may be justified 
in this sense [1-6]. To verify the validity of this expectation,
we investigate the $k_{max}$ dependence of a typical nucleon 
observable which contains only a logarithmic divergence, i.e.
the isovector axial-vector coupling constant $g_A^{(3)}$.
Fig.5 show the $k_{max}$ dependence of $g_A^{(3)}$ in the
single- and double-subtraction Pauli-Villars regularization 
schemes. One sees that this quantity certainly shows a tendency of
convergence in both regularization schemes, though the
rate of convergence in the double-subtraction scheme
is much faster than for the scalar and pseudoscalar densities in
the same regularization scheme. Nonetheless, one must be very careful if
one is interested in nucleon observables, which have
direct dependence on $S(r)$ or $P(r)$. The most important 
nucleon observable, which falls into this category, is the 
nucleon scalar charge (or the quark condensate in the nucleon)
given by
\begin{equation}
   \langle N | \bar{\psi} \psi | N \rangle \equiv 
   \int d^3 r \,[S(r) - S(r = \infty)]. 
\end{equation}
The superiority of the double-subtraction scheme to the
single-subtraction one must be self-explanatory in this case,
since this quantity is convergent only in the former scheme.

\subsection{Pauli-Villars versus proper-time regularization}

How to introduce ultraviolet cutoff into our effective chiral
theory is a highly nontrivial problem. Diakonov et al. advocated
the Pauli-Villars subtraction scheme as a ``good'' regularization scheme
for evaluating leading-twist parton distribution functions of the
nucleon within the chiral quark soliton model [1,2]. The reason is that
it preserves several general properties of the parton distributions
(such as positivity, factorization properties, sum rules etc.), which can
easily be violated by a naive ultraviolet regularization.
On the other hand, Schwinger's proper-time regularization has most
frequently been used for investigating low energy nucleon properties
within the chiral quark soliton model [7-9]. One might then wonder how
these predictions obtained by using the proper-time regularization scheme
would be altered if one uses the Pauli-Villars one.

Before entering into this discussion, we think it useful to recall
some basic properties of the proper-time regularization scheme. In this
scheme, the regularized effective meson action takes the same form
as (10) except that $S_f^{reg} [U]$ is now given in the form :
\begin{equation}
   S_f^{reg} [U] = \frac{1}{2} \,i \,N_c \int_0^\infty 
   \frac{d \tau}{\tau} \,\varphi (\tau) \,
   \mbox{Sp} \left( e^{- \tau D^\dagger D} - 
   e^{- \tau D_0^\dagger D_0} \right) ,
\end{equation}
with
\begin{equation}
   D = i \not\!\partial - M U^{\gamma_5} , \hspace{10mm}
   D_0 = i \not\!\partial - M .
\end{equation}
The regularization function $\varphi (\tau)$ is introduced so as to cut off
ultraviolet divergences which now
appear as a singularity at $\tau = 0$. For determining it, we can use
a similar criterion as what was used in the Pauli-Villars scheme.
That is, we require that the regularized theory reproduces the correct
normalization of the pion kinetic term as well as the empirical value
of the vacuum quark condensate. This gives two conditions :
\begin{eqnarray}
   \frac{N_c M^2}{4 \pi^2} \int_0^\infty \frac{d \tau}{\tau} \,
   \varphi (\tau) \,e^{- \tau M^2} &=& f_\pi^2 , \\
   \frac{N_c M}{2 \pi^2} \int_0^\infty \frac{d \tau}{\tau^2} \,
   \varphi (\tau) \,e^{- \tau M^2} &=& 
   {\langle \bar{\psi} \psi \rangle}_{vac} .
\end{eqnarray}
Schwinger's original choice corresponds to taking
\begin{equation}
   \varphi (\tau) = \theta \left( \tau - \frac{1}{\Lambda^2} \right) ,
\end{equation}
with $\Lambda$ being a physical cutoff energy. However, this simplest choice
cannot fulfill the two conditions (45) and (46) simultaneously.
Then, we use here slightly more complicated form as
\begin{equation}
   \varphi (\tau) = c \,\theta \left( \tau - \frac{1}{\Lambda_1^2} \right)
   + (1-c) \,\theta \left( \tau - \frac{1}{\Lambda_2^2} \right) ,
\end{equation}
which contains three parameters $c, \Lambda_1$ and $\Lambda_2$ [18].
Although the above two conditions are not enough to uniquely fix the
above three parameters, we find that solution sets
$(c, \Lambda_1, \Lambda_2)$ lie only in a small range of parameter
space and that this slight difference of regularization parameters
hardly affects the soliton properties. We use the following set of
parameters in the numerical investigation below :
\begin{equation}
   c = 0.720, \hspace{8mm} \Lambda_1 = 412.79 \,\mbox{MeV}, \hspace{8mm} 
   \Lambda_2 = 1330.60 \,\mbox{MeV}.
\end{equation}
Within the framework of the chiral quark soliton model, which assumes
slow collective rotation of a hedgehog soliton as
\begin{equation}
   U^{\gamma_5} (\mbox{\boldmath $x$},t) = A(t) U^{\gamma_5}_0 
   (\mbox{\boldmath $x$}) A^\dagger (t), \hspace{15mm}
   A(t) \subset \mbox{SU(2)} ,
\end{equation}
the nucleon matrix element of any quark bilinear operator $\bar{\psi}
O \psi$ is given as a perturbative series in the collective angular
velocity operator $\Omega$ defined by
\begin{equation}
   \Omega = i \,A^\dagger (t) \frac{d}{d t} A(t) .
\end{equation}
It is shown below that a noteworthy difference between the proper-time
regularization and the Pauli-Villars one appears at the zeroth order
term in $\Omega$. We recall that, in  both schemes,
the $O (\Omega^0)$ contribution to this matrix element is given as
\begin{equation}
   {\langle O \rangle}^{\Omega^0} = \int {\cal D} A \,\,
   \Psi_{M_J M_T}^{(J)^*} [A] \,
   {\langle O \rangle}_A^{\Omega^0} \,
   \Psi_{M_J M_T}^{(J)} [A] ,
\end{equation}
with
\begin{equation}
   {\langle O \rangle}^{\Omega^0}_A = {\langle O \rangle}_{val}^{\Omega^0}
    + {\langle O \rangle}_{vp}^{\Omega^0} ,
\end{equation}
where $\Psi_{M_J M_T}^{(J)} [A]$ is a wave function describing the
collective rotational motion. In eq.(53),
\begin{equation}
   {\langle O \rangle}_{val}^{\Omega^0} = 
   N_c \,\langle 0 | \tilde{O} | 0 \rangle , \ \ \ 
   \mbox{with} \ \ \ \ \tilde{O} = A^\dagger O A ,
\end{equation}
represents the contribution of the discrete
bound state level called the valence quark one.
Within the Pauli-Villars scheme, the contribution of the Dirac continuum
can be given in either of the following two forms :
\begin{eqnarray}
   {\langle O \rangle}_{vp}^{\Omega^0} &=& \ \,\,\,
   N_c \,\sum_{n < 0} \,\langle n | \tilde{O} | n \rangle - 
   \mbox{Pauli-Villars subtraction} , \nonumber \\
   &=& - \,N_c \,\sum_{n \ge 0} \,\langle n | \tilde{O} | n \rangle - 
   \mbox{Pauli-Villars subtraction} .
\end{eqnarray}
Note that the first form is given as a sum over the occupied single-quark
levels, while the second form given as a sum over the nonoccupied levels.
The equivalence of the two expressions follows from the
identity
\begin{equation}
   0 = \mbox{Sp} \,\tilde{O} = 
   \sum_{n < 0} \,\langle n | \tilde{O} | n \rangle + 
   \sum_{n \ge 0} \,\langle n | \tilde{O} | n \rangle ,
\end{equation}
which holds for most operators including the isovector magnetic
moment operator investigated below, if it is combined with the fact
that a similar identity holds also for the corresponding Pauli-Villars
subtraction terms. The situation is a little different for the proper-time
regularization scheme. The regularized Dirac sea contribution in this
scheme is given in the following form [8] :
\begin{equation}
   {\langle O \rangle}_{vp}^{\Omega^0} = - \frac{N_c}{2} \sum_{n = all}
   \mbox{sign} (E_n) g(E_n) \langle n | \tilde{O} | n \rangle ,
\end{equation}
with
\begin{equation}
   g(E_n) = \frac{1}{\sqrt{\pi}} \int_0^\infty 
   \frac{d \tau}{\sqrt{\tau}} \,| E_n | \,e^{- \tau E_n^2} .
\end{equation}
To compare this with the corresponding expression in the Pauli-Villars
scheme, it is convenient to rewrite it as
\begin{eqnarray}
   {\langle O \rangle}_{vp}^{\Omega^0} &=& \frac{1}{2} \,\Bigl\{ N_c 
   \sum_{n < 0} g(E_n) \langle n | \tilde{O} | n \rangle - 
   N_c \sum_{n \ge 0} g(E_n) \langle n | \tilde{O} | n \rangle  \Bigr\} .
\end{eqnarray}
One sees that here the answer is given as an average of the two
expressions, i.e. the one given as a sum over the occupied levels and
the others given as a sum over the nonoccupied levels.
(This feature is a consequence of the starting covariant expression for
an operator expectation value in the proper-time scheme.)
However, contrary to the previous case in which ultraviolet
regularization is introduced in the form of the Pauli-Villars
subtraction, now there is no reason to believe that the above two
terms give the same answer. In fact, the introduction of the
energy dependent cutoff factor $g(E_n)$ generally breaks the
equivalence of the two expressions because of the spectral asymmetry
of the positive- and negative-energy levels induced by the background
pion field of hedgehog form.

Now we start a comparative analysis of the two regularization schemes
on the basis of the numerical results. 
For reference, we also solve the soliton equation of motion
in the chiral limit. By assuming no (or at least weak) $m_\pi$
dependence of ${< \bar{\psi} \psi >}_{vac}$ appearing in (16) and (38),
this calculation can be done by setting $m_\pi = 0$ in (18) and (25)
without changing the sets of regularization parameters given in (41)
and (49). Since the way of cutting off
the ultraviolet component is totally different for the two regularization
schemes, it naturally affects solutions
of the soliton equation of motion. Although the detailed contents of the
soliton energy are highly model dependent concepts and are not direct
observables, they are anyhow very sensitive to this difference of the
self-consistent solutions. Table 1 shows this comparison.
Comparing the answers of the two regularization schemes, one finds that
the Pauli-Villars scheme leads to more strongly deformed soliton, which
means a deeper binding of the discrete valence level and larger vacuum
polarization energy. One sees that the total soliton energy is lower
for the Pauli-Villars scheme than for the proper-time scheme.
One also observes that the soliton energy is very sensitive to the
pion mass. When one goes from the finite pion mass case to the chiral
limit, one obtains much lower soliton energy.

\begin{table}[h]
\begin{center}
\renewcommand{\baselinestretch}{1.2}
\caption{The static soliton energy in the proper-time regularization
schme and the (double-subtraction) Pauli-Villars one.
$E_{val}, E_{v.p.}^{reg}$ respectively stand for the valence quark
contribution and the Dirac sea one to the fermionic
energy, while $E_m$ represents the mesonic part of the energy.
The sum of these three parts gives the total static energy
$E_{static}^{reg}$.}
\renewcommand{\baselinestretch}{1.38}
\vspace{5mm}
\begin{tabular}{ccccc}
\hline\hline
 & $E_{val}$ [MeV] & $E_{v.p.}^{reg}$ [MeV] & $E_m$ [MeV] & $E_{static}^{reg}$
 [MeV] \\
\hline
proper-time ($m_\pi = 138 \,\mbox{MeV}$) & 633.0 & 617.6 & 37.2 & 1287.9 \\
Pauli-Villars ($m_\pi = 138 \,\mbox{MeV}$) & 447.6 & 569.2 & 51.3 & 1068.1 \\
\hline
proper-time ($m_\pi = 0 \,\mbox{MeV}$) & 555.6 & 688.6 & 0 & 1244.2 \\
Pauli-Villars ($m_\pi = 0 \,\mbox{MeV}$) & 351.5 & 655.4 & 0 & 1006.9 \\
\hline\hline
\end{tabular}
\end{center}
\end{table}

\begin{table}[h]
\begin{center}
\renewcommand{\baselinestretch}{1.2}
\caption{The quark spin content of the nucleon $< \Sigma_3 >$
in the proper-time regularization scheme and the Pauli-Villars
one.}
\renewcommand{\baselinestretch}{1.38}
\vspace{5mm}
\begin{tabular}{cccc}
\hline\hline
 & ${< \Sigma_3 >}_{val}$ & ${< \Sigma_3 >}_{v.p.}$ & ${< \Sigma_3 >}$ \\
\hline
proper-time ($m_\pi = 138 \,\mbox{MeV}$) & 0.484 & 0.005 & 0.489 \\
Pauli-Villars ($m_\pi = 138 \,\mbox{MeV}$) & 0.391 & 0.008 & 0.399 \\
\hline
proper-time ($m_\pi = 0 \,\mbox{MeV}$) & 0.374 & 0.007 & 0.380 \\
Pauli-Villars ($m_\pi = 0 \,\mbox{MeV}$) & 0.286 & 0.011 & 0.298 \\
\hline\hline
\end{tabular}
\end{center}
\end{table}

Probably, the most important observable which has strong sensitivity to the
above difference of the self-consistent solutions is the flavor-singlet
axial charge or the quark spin content of the nucleon ${\langle \Sigma_3
\rangle}$. The theoretical predictions for this quantities in the
two regularization schemes are shown in Table 2. In evaluating this
quantity, we did not introduce any regularization, because it is
related to the imaginary part of the (Euclidean) effective action and
is convergent itself. This means that the difference between the two
schemes purely comes from that of the self-consistent solutions.
One sees that the Pauli-Villars scheme leads to smaller quark spin
content. The reason can easily be understood. Within the framework of
the chiral quark soliton model, the rest of the nucleon spin is
carried by the orbital angular momentum of quark fields and this
latter portion increases as the deformation of the soliton becomes
larger [8]. A similar tendency is also observed when one goes from
the finite pion mass case to the chiral limit.

\begin{table}[h]
\begin{center}
\renewcommand{\baselinestretch}{1.2}
\caption{The $O (\Omega^0)$ contributions to the
isovector magnetic moment of the nucleon in the proper-time
regularization scheme and the Pauli-Villars one.
The second column represents for the valence quark contribution.
The third and fourth columns stand for the answers
for the vacuum polarization contributions respectively obtained
with the occupied and nonoccupied formulas, while the fifth column
gives the average of the two answers. The total $O(\Omega^0)$
contributions are shown in the sixth column.}
\renewcommand{\baselinestretch}{1.38}
\vspace{5mm}
\begin{tabular}{cccccc}
\hline\hline
 & \raisebox{-1.5ex}[0pt]{$\mu_{val}^{(3)} (\Omega^0)$}
 & \multicolumn{1}{c}{} & 
 \multicolumn{1}{c}{\raisebox{-1mm}[0pt]{$\mu_{v.p.}^{(3)reg} (\Omega^0)$}} &
 \multicolumn{1}{c}{} & 
 \raisebox{-1.5ex}[0pt]{$\mu^{(3)} (\Omega^0)$} \\
 &  & \raisebox{1mm}[0pt]{occupid} & 
 \raisebox{1mm}[0pt]{nonoccupied} & \raisebox{1mm}[0pt]{average} &  \\
\hline
proper-time ($m_\pi = 138 \,\mbox{MeV}$) & 1.611 & 1.312 & 0.210 & 0.761 & 2.372 \\
Pauli-Villars ($m_\pi = 138 \,\mbox{MeV}$) & 1.762 & 0.996 & 0.996 & 0.996 & 2.759 \\
\hline
proper-time ($m_\pi = 0 \,\mbox{MeV}$) & 1.623 & 1.908 & 0.588 & 1.248 & 2.875 \\
Pauli-Villars ($m_\pi = 0 \,\mbox{MeV}$) & 1.810 & 1.738 & 1.738 & 1.738 & 3.547 \\
\hline\hline
\end{tabular}
\end{center}
\end{table}

There are different kinds of nucleon observables, which contain (potential)
logarithmic divergence and thus depend directly on how they are regularized.
Most typical are the $O (\Omega^0)$ contribution to the isovector
axial-vector coupling constant $g_A^{(3)}$ and the isovector
magnetic moment $\mu^{(3)}$ of the nucleon. Let us first show the results for
the isovector magnetic moment, since it turns out to
have stronger dependence on the choice of the regularization scheme.
Table 3 shows the $O (\Omega^0)$ contribution to the isovector magnetic
moment. For each regularization scheme, the second column represents
the answer obtained with the occupied expression, while the third
column does the answer obtained with the nonoccupied one. In the case of
Pauli-Villars scheme, the equivalence of the two expressions is
nicely confirmed by the explicit numerical calculation.
In the case of proper-time scheme, however, we encounter quite a
dissimilar situation. First, the answer obtained with the occupied
expression is about $30 \,\%$ larger than the corresponding answer
of the Pauli-Villars scheme, while the answer obtained with the
nonoccupied expression is about $80 \,\%$ smaller than the answer obtained
with the occupied one. Since the final answer of the proper-time scheme
is given as an average of the occupied and nonoccupied expressions,
the consequence is that the
prediction of the proper-time scheme for the $O (\Omega^0)$
contribution to $\mu^{(3)}$ is about $14 \,\%$ smaller than the corresponding
prediction of the Pauli-Villars scheme. (See the fourth column of the
Table 3.) Note that the difference between the two regularization schemes
becomes much more drastic when one goes to the chiral limit. This is
due to the fact that the $O (\Omega^0)$ vacuum polarization contribution to
the isovector magnetic moment is extremely sensitive to the pion mass
effect such that it is much larger in the chiral limit.

\begin{table}[h]
\begin{center}
\renewcommand{\baselinestretch}{1.2}
\caption{The final predictions for the isovector magnetic moment of the
nucleon, given as sums of the $O(\Omega^0)$ and
$O(\Omega^1)$ contributions.}
\renewcommand{\baselinestretch}{1.38}
\vspace{5mm}
\begin{tabular}{cccc}
\hline\hline
 & $\mu^{(3)} (\Omega^0)$ & $\mu^{(3)} (\Omega^1)$ & $\mu^{(3)} (\Omega^0 + \Omega^1)$ \\
\hline
proper-time ($m_\pi = 138 \,\mbox{MeV}$) & 2.372 & 1.072 & 3.445 \\
Pauli-Villars ($m_\pi = 138 \,\mbox{MeV}$) & 2.759 & 1.211 & 3.970 \\
\hline
proper-time ($m_\pi = 0 \,\mbox{MeV}$) & 2.875 & 1.032 & 3.907 \\
Pauli-Villars ($m_\pi = 0 \,\mbox{MeV}$) & 3.547 & 1.182 & 4.729 \\
\hline\hline
\end{tabular}
\end{center}
\end{table}

\begin{table}[h]
\begin{center}
\renewcommand{\baselinestretch}{1.2}
\caption{The final predictions for the isovector axial-coupling
constant of the nucleon, given as sums of the $O(\Omega^0)$ and
$O(\Omega^1)$ contributions.}
\renewcommand{\baselinestretch}{1.38}
\vspace{5mm}
\begin{tabular}{cccc}
\hline\hline
 & $g_A^{(3)} (\Omega^0)$ & $g_A^{(3)} (\Omega^1)$ & $g_A^{(3)} (\Omega^0 + \Omega^1)$ \\
\hline
proper-time ($m_\pi = 138 \,\mbox{MeV}$) & 0.848 & 0.412 & 1.260 \\
Pauli-Villars ($m_\pi = 138 \,\mbox{MeV}$) & 0.976 & 0.408 & 1.384 \\
\hline
proper-time ($m_\pi = 0 \,\mbox{MeV}$) & 0.921 & 0.348 & 1.269 \\
Pauli-Villars ($m_\pi = 0 \,\mbox{MeV}$) & 1.054 & 0.344 & 1.398 \\
\hline\hline
\end{tabular}
\end{center}
\end{table}

Before comparing our theoretical predictions with the observed
isovector magnetic moment of the nucleon, we must take account of
the $O (\Omega^1)$ contribution, too, since it is known to give
sizable correction to the leading-order result [19,20]. Although we
do not go into the detail here, it turns out that this $O (\Omega^1)$
piece is not so sensitive to the difference of the regularization
scheme as the $O (\Omega^0)$ piece is. The reason is that this
$O (\Omega^1)$ term is given as a double sum over the occupied
levels and the nonoccupied ones and the formula has some symmetry
under the exchange of these two types of single-quark orbitals [21].
The final predictions for the nucleon isovector magnetic moment
obtained as a sum of the $O (\Omega^0)$ and $O (\Omega^1)$
contributions are shown in Table 4. After all, the prediction
of the Pauli-Villars scheme is about $15 \,\%$ larger than that of the
proper-time scheme and a little closer to the observed moment.
The effect is much more drastic in the chiral limit. The prediction
of the Pauli-Villars scheme is about $20 \,\%$ larger than that of
the proper-time scheme and nearly reproduces the observed
isovector magnetic moment of the nucleon, i.e.
$\mu_{exp}^{(3)} \simeq 4.71$.

Finally, we show in Table 5 the predictions for the isovector
axial-charge of the nucleon obtained as a sum of the $O (\Omega^0)$
and $O (\Omega^1)$ contributions. Also for this quantity, there are
some detailed differences between the predictions of the two
regularization schemes. Nonetheless, the final answers for
$g_A^{(3)}$ turn out to be not so sensitive to the difference of
the regularization schemes as compared with the case of the
isovector magnetic moment. Besides, one also notices that the
finite pion mass effect hardly influences the final prediction
for this particular quantity.

\section{Conclusion}

In summary, the single-subtraction Pauli-Villars regularization
scheme, which is often used in evaluating nucleon structure functions
within the framework of the CQSM, cannot be regarded as
a fully consistent regularization scheme in that it still contains
ultraviolet divergences in the scalar and psuedoscalar quark densities
appearing in the soliton equation of motion.
However, these divergences can easily be removed by
increasing the number of subtraction term from one to two.
After this straightforward generalization, the effective theory is
totally divergence free. Especially, both the vacuum quark condensate
and the isoscalar piece of the nucleon scalar charge becomes finite now.
Nonetheless, we find that, owing to the accidental
cancellation explained in the text, one can obtain a finite soliton
profile $F(r)$ even in the single-subtraction scheme, and besides the
resultant soliton solution is not extreme different from the
corresponding one obtained in the double-subtraction scheme.
Furthermore, it turns out that, for most nucleon observables, which
contain only the logarithmic divergence, the predictions of the two
regularization schemes are not much different. The previous calculations
of quark distribution functions with use of the single-subtraction
Pauli-Villars regularization scheme would be justified in this sense.

We have also carried out a comparative analysis of typical nucleon
observables based on the Pauli-Villars regularization scheme and the
proper-time one. A nice property of the Pauli-Villars regularization
scheme, which is not possessed by the proper-time one, is that it
preserves a nontrivial symmetry of the original theory, i.e. the
equivalence of the occupied and nonoccupied expressions for
$O (\Omega^0)$ contributions to nucleon observables. The improvement
obtained for the isovector magnetic moment of the nucleon was shown
to be related to this favorable property of the Pauli-Villars
regularization scheme. How to introduce ultraviolet cutoff into an
effective low energy model should in principle be predictable from
the underlying QCD dynamics. For lack of precise information about it,
however, phenomenology must provides us with an important criterion
for selecting regularization schemes. The regularization scheme
based on the Pauli-Villars subtraction appears to be a good candidate
also in this respect.

\vspace{3mm}
\section*{Acknowledgement}

Numerical calculation was performed by using the workstations
at the Laboratory of Nuclear Studies, and those at the Research Center
for Nuclear Physics, Osaka University. 

%
%
\vspace{3mm}
\section*{References}
\newcounter{refnum}
\begin{list}%
{[\arabic{refnum}]}{\usecounter{refnum}}
\item D.I.~Diakonov, V.Yu.~Petrov, P.V.~Pobylitsa, M.V.~Polyakov,
and C.~Weiss, \\
Nucl. Phys. {\bf B480}, 341 (1996).
\item D.I.~Diakonov, V.Yu.~Petrov, P.V.~Pobylitsa, M.V.~Polyakov,
and C.~Weiss, \\
Phys. Rev. {\bf D56}, 4069 (1997).
\item M.~Wakamatsu and T.~Kubota, Phys. Rev. {\bf D56}, 4069 (1998).
\item M.~Wakamatsu and T.~Kubota, Osaka University preprint
OU-HET-310/98, \\
hep-ph/9809443.
\item C.~Weiss and K.~Goeke, Bochum University preprint RUB-TPII-12/97,\\
hep-ph/9712447.
\item P.V.~Pobylitsa, M.V.~Polyakov, K.~Goeke, T.~Watabe, and C.~Weiss,\\
Bochum University preprint RUB-TPII-4/98, hep-ph/9804436.
\item H.~Reinhardt and R.~W\"{u}nsch, Phys. Lett. {\bf B215},
577 (1998) ;\\
T.~Meissner, F.~Gr\"{u}mmer, and K.~Goeke, Phys. Lett. {\bf B227},
296 (1989).
\item M.~Wakamatsu and H.~Yoshiki, Nucl. Phys. {\bf A524}, 561 (1991).
\item For reviews, see, M.~Wakamatsu, Prog. Theor. Phys. Suppl.
{\bf 109}, 115 (1992) ; \\
Chr.V.~Christov, A.~Blotz, H.-C.~Kim, P.~Pobylitsa, T.~Watabe,
Th.~Meissner, \\
E.~Ruiz Arriola and K.~Goeke, Prog. Part.
Nucl. Phys. {\bf 37}, 91 (1996) ;\\
R.~Alkofer, H.Reinhardt and H.~Weigel, Phys. Rep. {\bf 265}, 139
(1996).
\item H.~Weigel, L.~Gamberg, and H.~Reinhardt, Phys. Rev. {\bf D58},
038501 (1998).
\item D.I.~Diakonov, V.Yu.~Petrov, P.V.~Pobylitsa, M.V.~Polyakov,
and C.~Weiss, \\
Phys. Rev. {\bf D58}, 038502 (1998).
\item F.~D\"{o}ring, A.~Blotz, C.~Sch\"{u}ren, Th.~Meissner,
E.~Ruiz-Arriola, and K.~Goeke, \\
Nucl. Phys. {\bf A536}, 548 (1992).
\item T.~Watabe and H.~Toki, Prog. Theor. Phys. {\bf 87}, 651 (1992).
\item M.~Jaminon, G.~Ripka, and P.~Stassart, Phys. Lett. {\bf B227},
191 (1989).
\item M.~Wakamatsu, Phys. Rev. {\bf D46}, 3762 (1992).
\item D.I.~Diakonov, V.Yu.~Petrov, and P.V.~Pobylitsa, Nucl. Phys.
{\bf B306}, 809 (1988).
\item S.~Kahana and G.~Ripka, Nucl. Phys. {\bf A429}, 462 (1984) ;\\
S.~Kahanna, G.~Ripka, and V.~Soni, Nucl. Phys. {\bf A415}, 351 (1984).
\item A.~Blotz, D.I.~Diakonov, K.~Goeke, N.W.~Park, V.Yu.~Petrov,
and P.V.~Pobylitsa, \\
Nucl. Phys. {\bf A555}, 765 (1993).
\item M.~Wakamatsu and T.~Watabe, Phys. Lett. {\bf B312}, 184 (1993).
\item Chr.V.~Christov, A.~Blotz, K.~Goeke, P.V.~Pobylitsa,
V.Yu.~Petrov, \\
M.~Wakamatsu, and T.~Watabe, Phys. Lett. {\bf B325},
467 (1994).
\item M.~Wakamatsu, Prog. Theor. Phys. {\bf 95}, 143 (1996).
\end{list}

\vspace{8mm}
\begin{flushleft}
\Large\bf{Figure caption} \\
\end{flushleft}
\ \\
\begin{minipage}{2cm}
Fig. 1.
\end{minipage}
\begin{minipage}[t]{13cm}
The $k_{max}$ dependence of the scalar quark density $S(r)$
and the pseudoscalar density $P(r)$ in the single-subtraction
Pauli-Villars scheme.
\end{minipage}
\ \\
\vspace{6mm}
\ \\
\begin{minipage}{2cm}
Fig. 2.
\end{minipage}
\begin{minipage}[t]{13cm}
The $k_{max}$ dependence of the scalar quark density $S(r)$
and the pseudoscalar density $P(r)$ in the double-subtraction
Pauli-Villars scheme.
\end{minipage}
\ \\
\vspace{6mm}
\ \\
\begin{minipage}{2cm}
Fig. 3.
\end{minipage}
\begin{minipage}[t]{13cm}
The scalar quark densities at the spatial infinity $S(r = \infty)$
as functions of $k_{max} / M$ and as functions of
$\log (k_{max} / M)$ in the single- and double-subtraction
Pauli-Villars schemes.
\end{minipage}
\ \\
\vspace{6mm}
\ \\
\begin{minipage}{2cm}
Fig. 4.
\end{minipage}
\begin{minipage}[t]{13cm}
The $k_{max}$ dependence of the self-consistent soliton profiles
$F(r)$ in the single- and double-subtraction Pauli-Villars
schemes. The curves with different $k_{max}$ are almost
indistinguishable.
\end{minipage}
\ \\
\vspace{6mm}
\ \\
\begin{minipage}{2cm}
Fig. 5.
\end{minipage}
\begin{minipage}[t]{13cm}
The $k_{max}$ dependence of the nucleon isovector axial-charges
$g_A^{(3)}$ in the single- and double-subtraction Pauli-Villars
schemes.
\end{minipage}
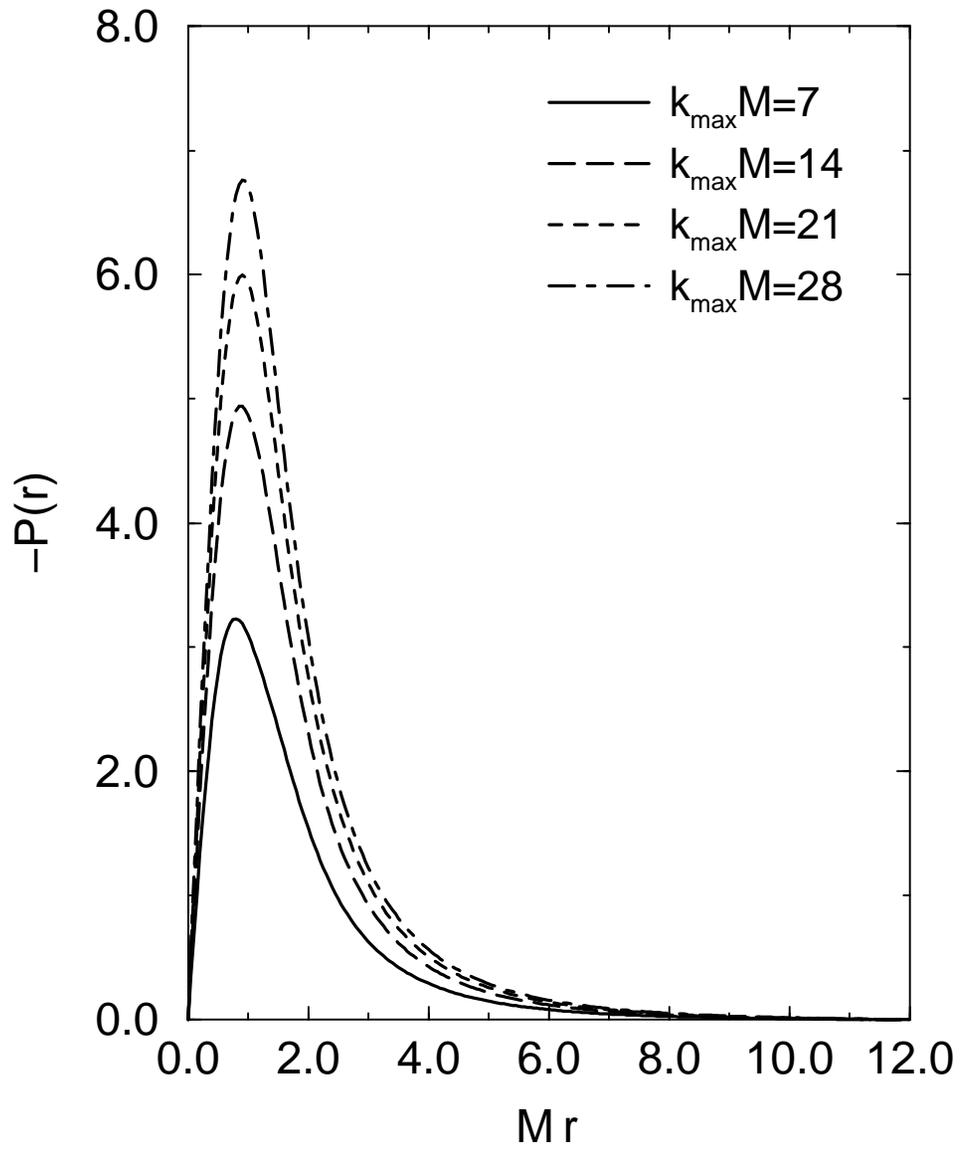
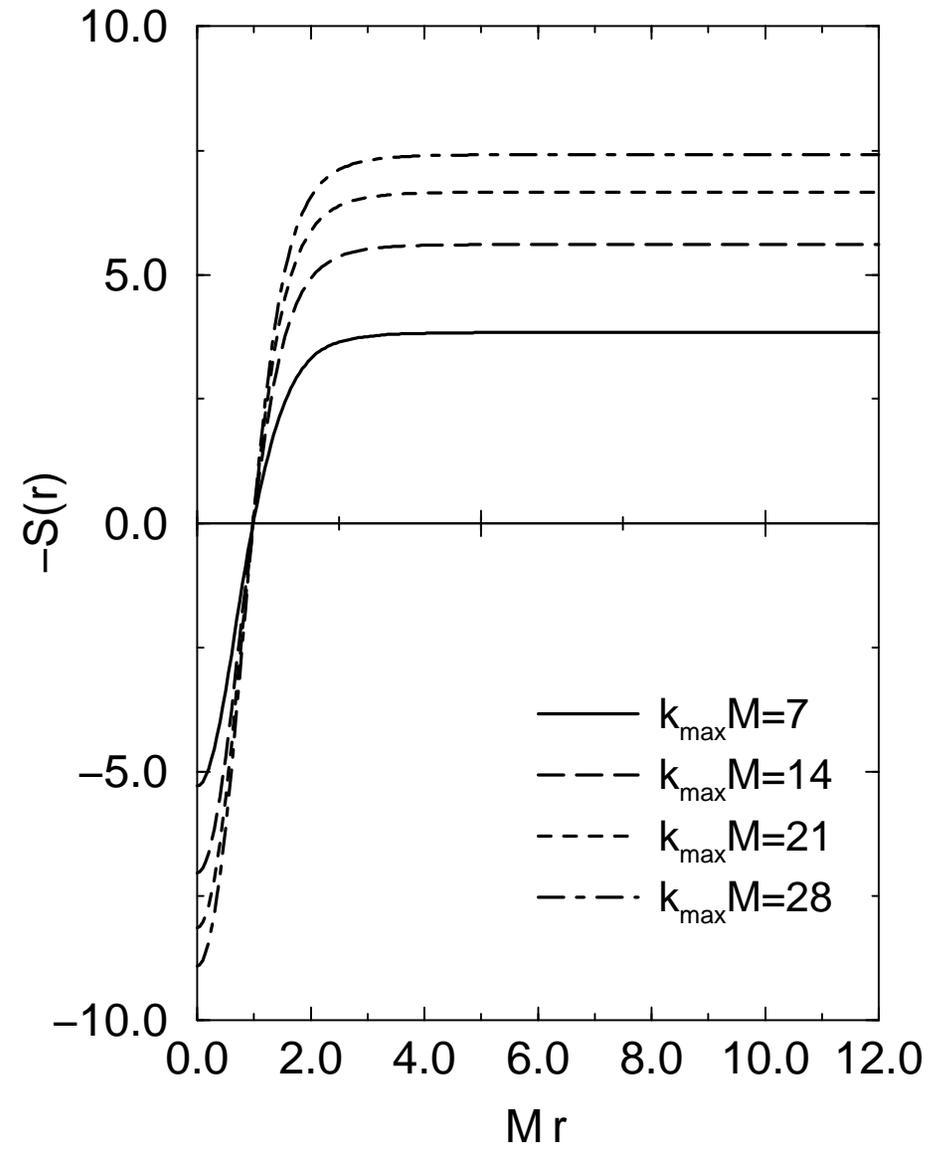
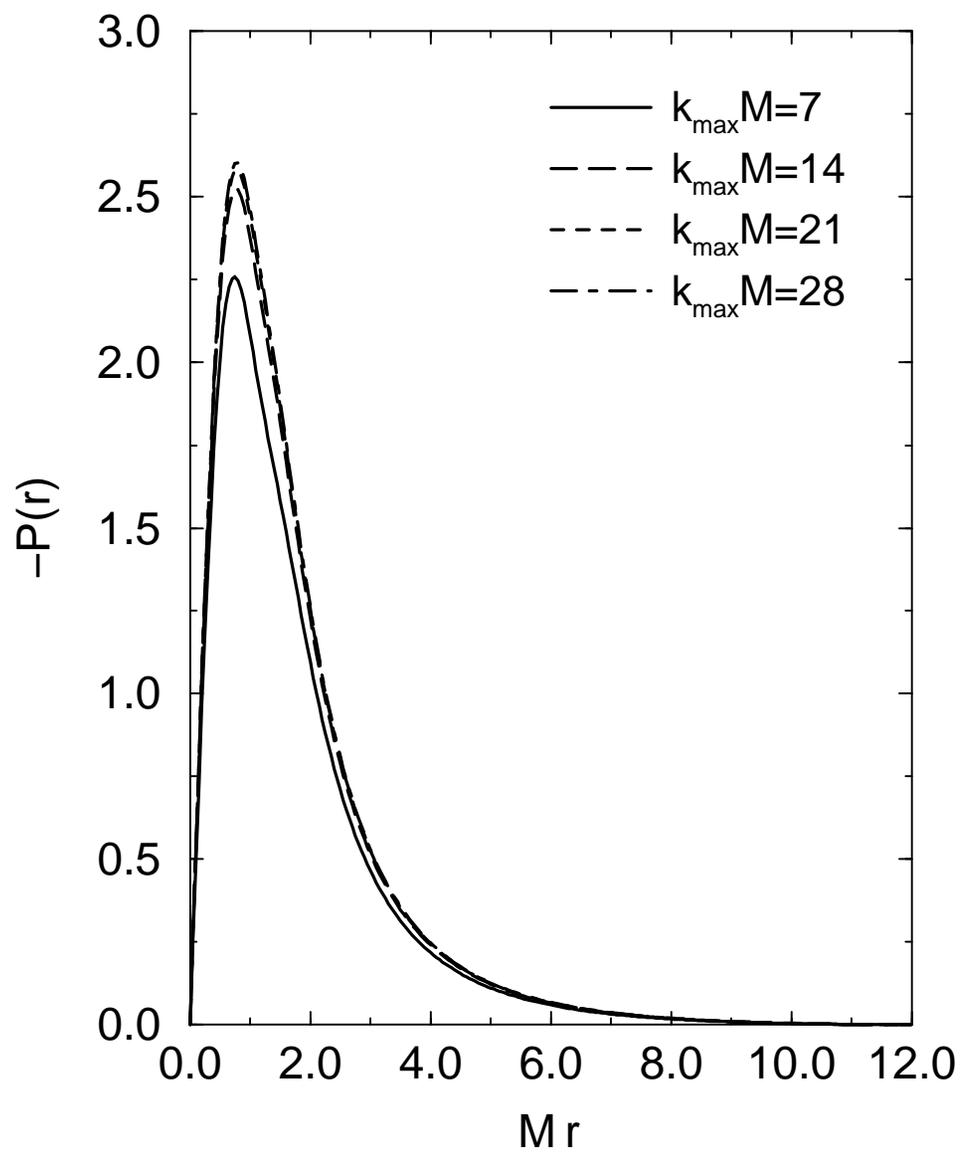
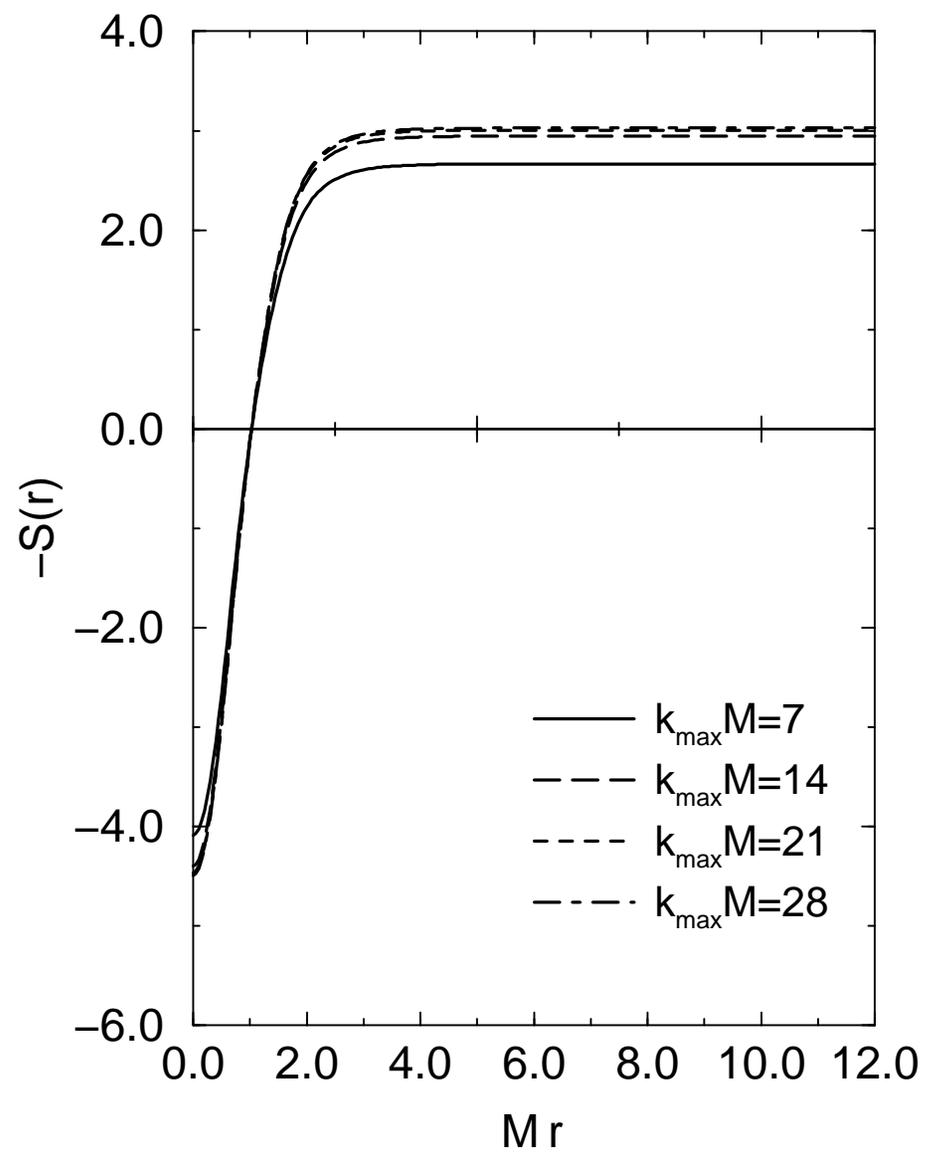
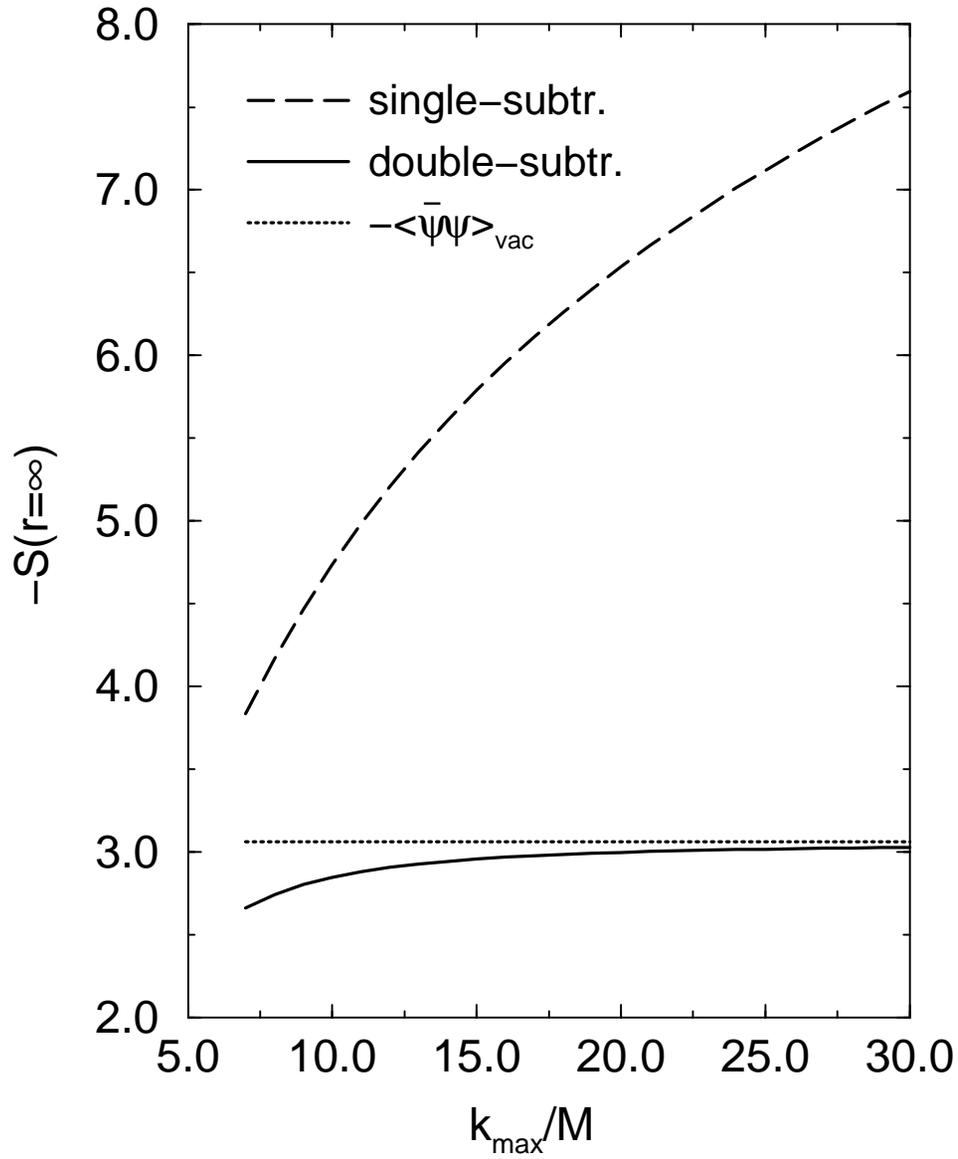
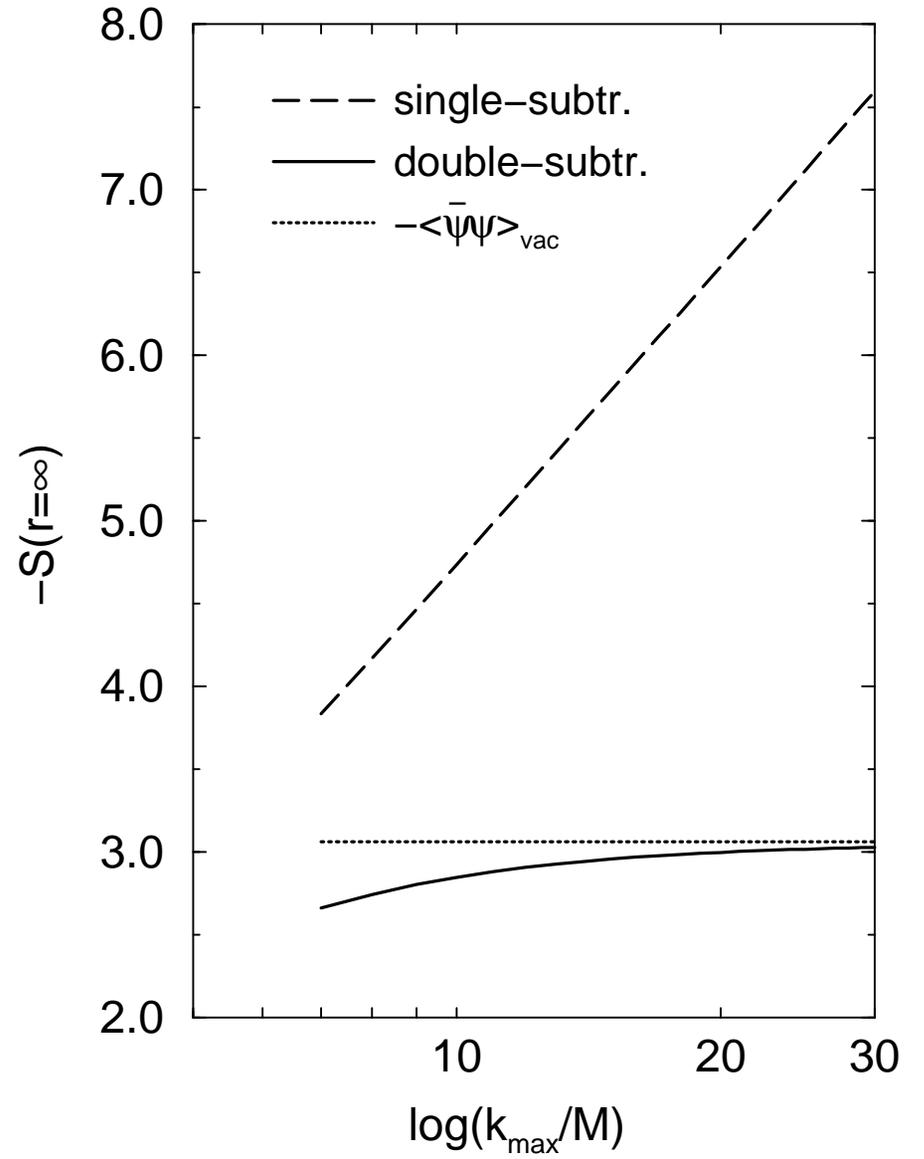
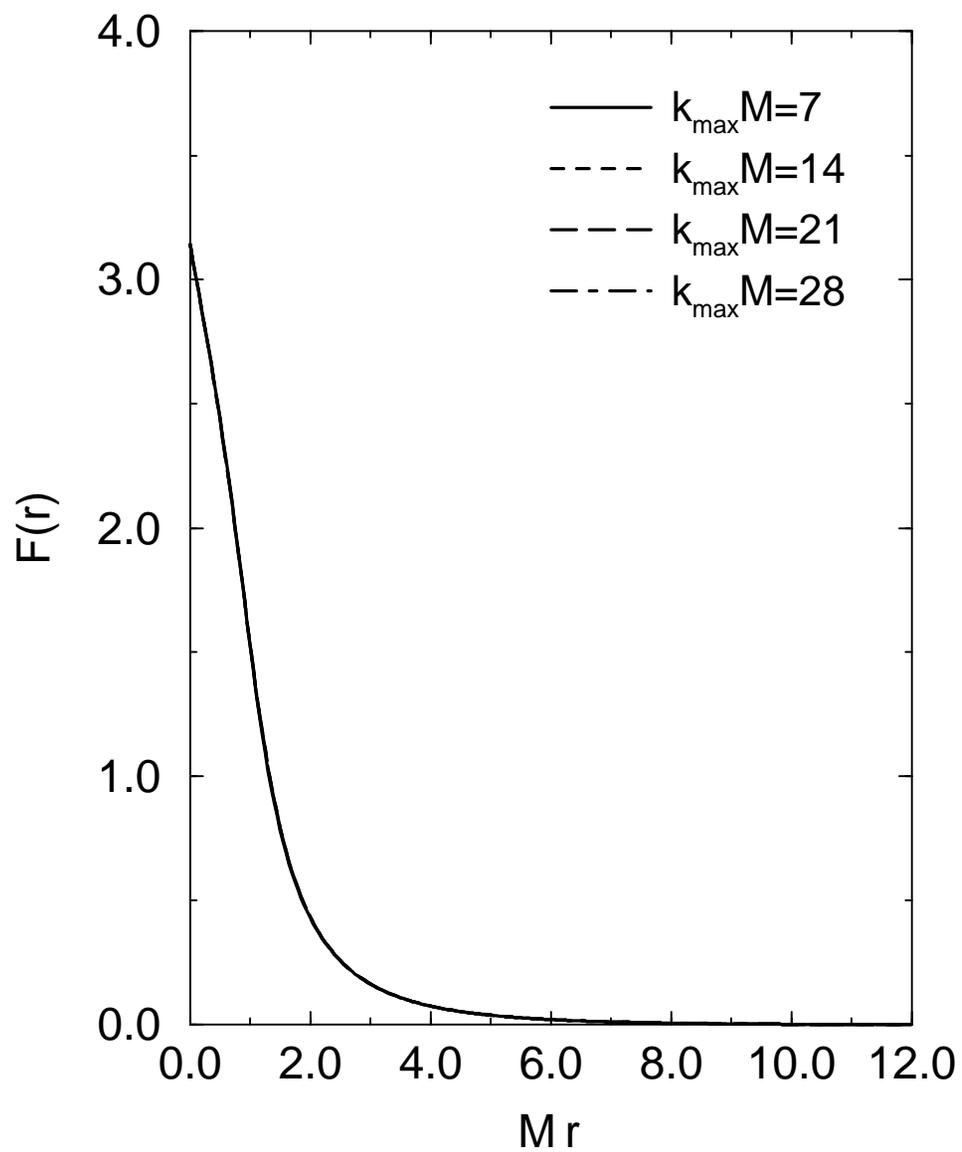
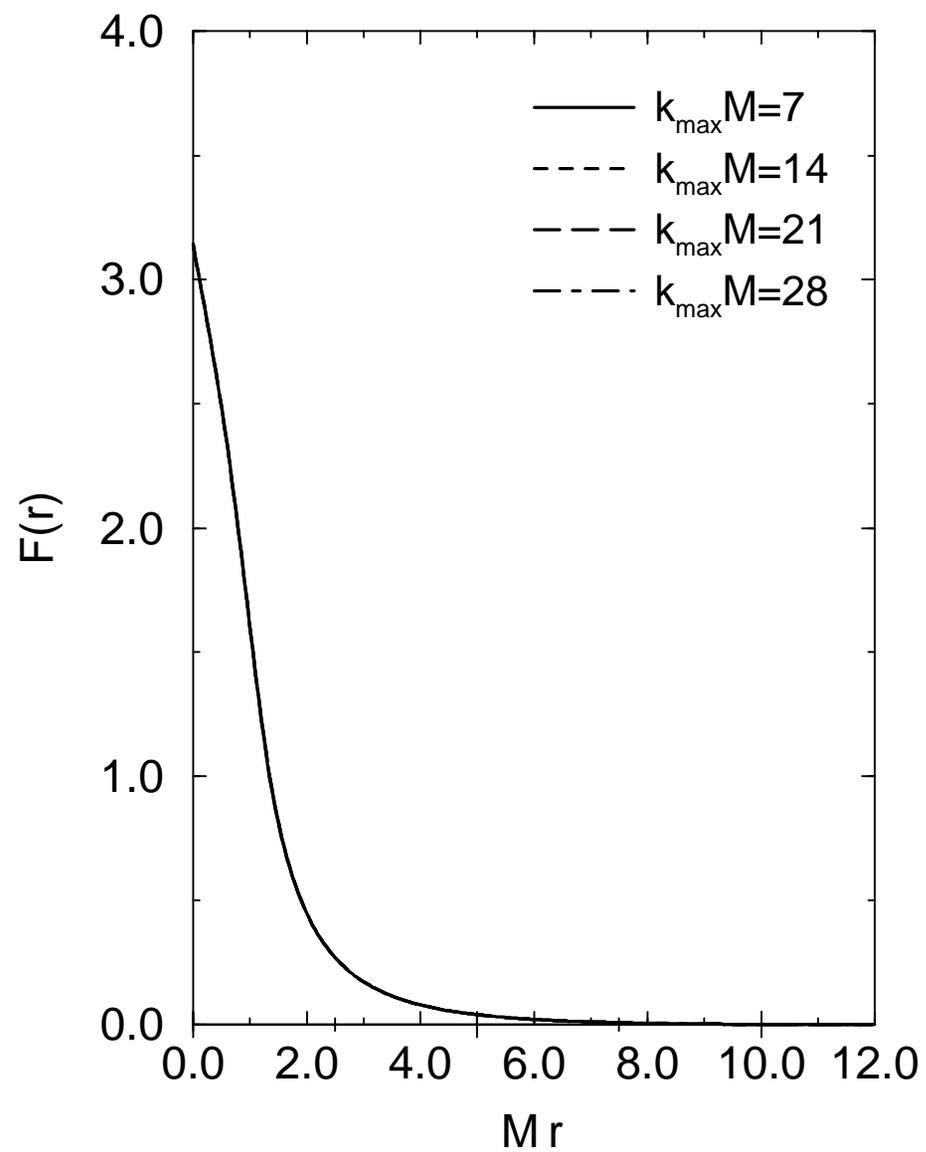
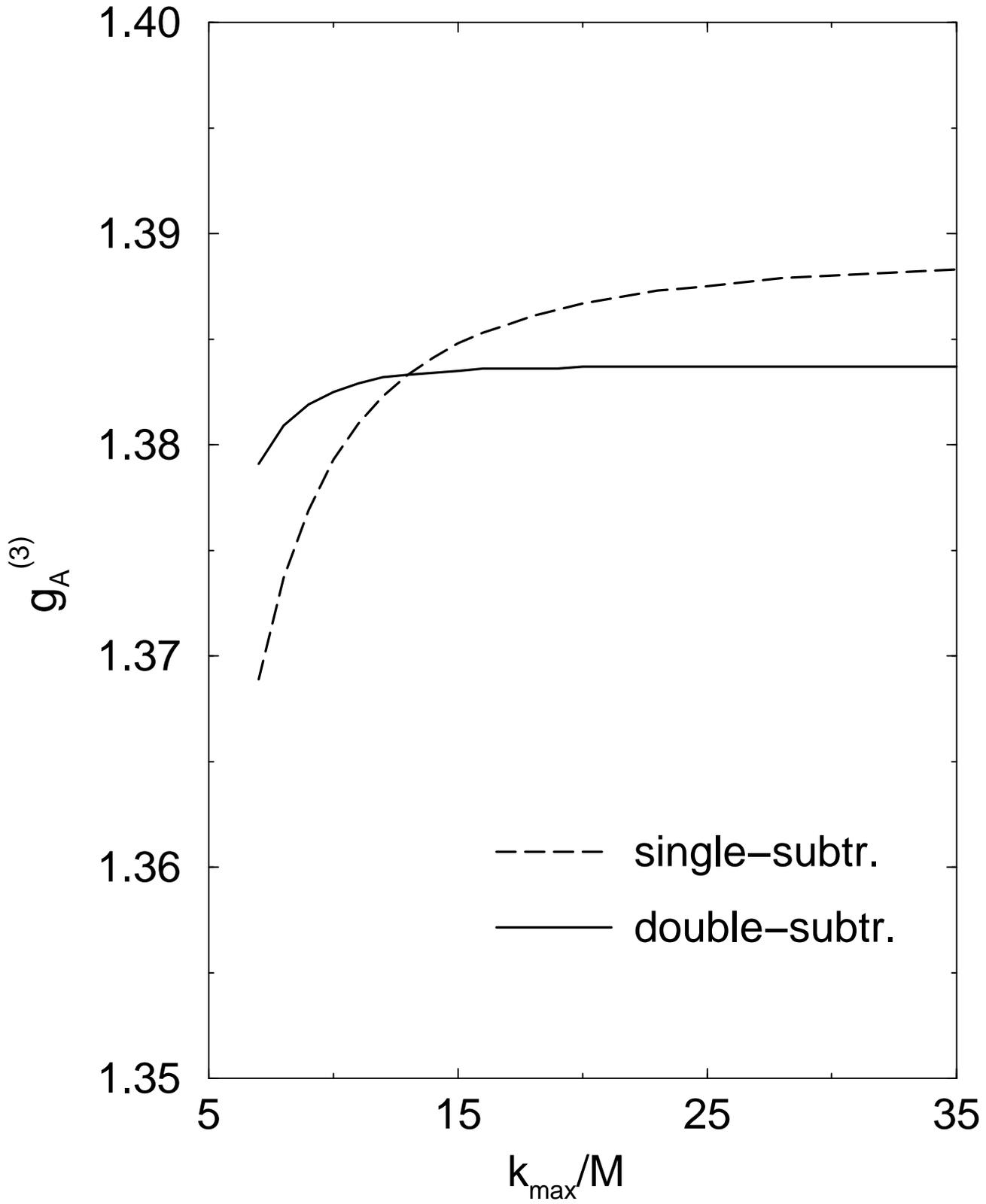
\end{document}